\documentclass[prl,showpacs,noshowkeys,amsmath,superscriptaddress,twocolumn,
a4paper]{revtex4}
\usepackage{ae,aecompl}
\usepackage{longtable}
\usepackage{graphicx}
\usepackage{amsmath}
\usepackage{amssymb}
\usepackage{amscd}

\def\*{******************************************}

\begin{document}

\title{Supersymmetry and Nonequilibrium Work Relations}
\author{Kirone Mallick }
 \affiliation{Service de Physique Th\'eorique, Centre d'Etudes de Saclay,
 91191 Gif-sur-Yvette Cedex, France}
 \author{Moshe Moshe}
\affiliation{Department of Physics,
  Technion - Israel Institute of Technology, Haifa ~32000, Israel}
  \author{Henri Orland}
 \affiliation{Service de Physique Th\'eorique, Centre d'\'Etudes de Saclay,
 91191 Gif-sur-Yvette Cedex, France}



\date{\today}
\begin{abstract}
  We give a field-theoretic proof of the nonequilibrium work relations   for
  a space-dependent field with stochastic dynamics. The path
  integral representation and its symmetries
 allow us to derive Jarzynski's equality. In addition, we derive a set of exact identities
 that generalize the fluctuation-dissipation relations to far-from-equilibrium situations. These identities  are prone to experimental verification.
 Furthermore, we show that  supersymmetry  invariance  of the
  Langevin equation, which is broken when the external  potential is time-dependent,
  is partially restored by adding to the action  a term which is precisely
  Jarzynski's  work. Jarzynski's equality can also be deduced from
 this supersymmetry.
\end{abstract}
 \maketitle


  During the last decade a number of exact relations have been derived
 for non-equilibrium processes. The
 Jarzynski equality is one of these remarkable results:
  it  shows that the statistical properties of the
  work performed on a system in contact with
 a heat reservoir at temperature $kT = \beta^{-1}$
 during a {\it  non-equilibrium}  process are
  related  to the free energy difference $\Delta F$
   between two  {\it equilibrium}  states of that system.
   This identity  was  derived originally using a Hamiltonian
 formulation \cite{jarzynskiPRL} and was  extended to
 systems obeying  a
 Langevin equation \cite{jarzynskiPRE} or a discrete  Markov equation
   \cite{crooks1,crooks2}.  Jarzynski's result
 has been verified
  on exactly solvable models \cite{jarmazonka}
  and  by explicit  calculations in
 kinetic theory of gases \cite{lua,vandenbroeck}. This  equality
 has also  been used  in   various
 single-molecule pulling experiments
 \cite{hummer,liphardt,ritort} to measure folding free energies
 and has been checked against analytical predictions
 on  mesoscopic mechanical devices
 such as a torsion pendulum \cite{douarche}.
  These  experiments   are delicate  to carry out  because
 the  mathematical validity of Jarzynski's theorem
    is insured by rare events that occur  with a  probability
  that typically decreases  exponentially with the system size
  (for a review see e.g. \cite{ritortrev}).

 In the present work, we derive  nonequilibrium
 work relations   for a field $\phi(x,t)$
 representing  a  coarse-grained  order parameter  of a microscopic system.
  This field  evolves according to an
 effective stochastic equation
  that depends  on the  symmetries and conservation
 laws of the system  \cite{hohenberg}. We represent
  the stochastic  evolution as
   a path integral  \cite{abishek1,jarzynski4,chetrite} and
 use   the  response field formalism \cite{msr}
 to  derive the  work relations  from the
invariance of the path integral under  certain  changes of variables.
We  obtain correlator identities that generalize the fluctuation-dissipation relations arbitrarily far from equilibrium. These identities can be checked experimentally in single molecule experiments.
 Furthermore,  by introducing auxiliary Grassmannian  fields,
 we interpret   this  invariance   as a manifestation of a  hidden
 supersymmetry.
  This   supersymmetry is known
 to be the fundamental  invariance  property   that
  embodies the
  principle of microscopic
 reversibility and leads  to   the fluctuation dissipation-theorem
 and the  Onsager reciprocity relations for a system
 at thermal equilibrium  \cite{tsvelik,gozzi,chaturvedi,JZJ}.
  Here, we show that,  far from  equilibrium,
  Jarzynski's  theorem, is
 also a  consequence of an  underlying  supersymmetry. This supersymmetry in turn allows to generalize the  fluctuation dissipation-theorem to far from equilibrium situations.

\noindent We   consider  a scalar  field  $\phi(x,t)$
  defined on  d-dimensional space with   Model A dynamics
  that  describes
  a system with non-conserved  order-parameter  \cite{hohenberg}
  ({\it e.g.},  the Ising model
 with Glauber dynamics):

\begin{equation}
    \frac{ \partial \phi}{\partial t}
   =
    - \Gamma_0 \frac {\delta {\mathcal U}}{\delta  \phi} +   \zeta(x,t)
     = - \Gamma_0 f(\phi) +   \zeta(x,t)
 \label{modelA}
\end{equation}
\noindent $\zeta(x,t)$ is a Gaussian
 white noise of zero mean value and correlations
 $
 \langle  \zeta(x,t) \zeta(x',t') \rangle = 2  \Gamma_0  kT
  \delta(t -t')\delta^d(x -x')$, \noindent $T$ being  the temperature. The dynamics is  thus
governed by the time-dependent
 potential
\begin{equation}
{\mathcal U}[\phi(x,t),t]
  = {\mathcal F}[\phi(x,t)] - \int d^dx \ h(x,t) \phi(x,t)
\label{def:Ut} \end{equation}

 \noindent The potential energy  (or Euclidian action)
  ${\mathcal F}[\phi]$ is, for instance, given by
  \begin{equation}
 {\mathcal F}[\phi] =
  \int {\rm d}^d x \{ \frac{1}{2} r_0  \phi^2 + \frac{1}{2}
  |\nabla\phi|^2  + u_0 \phi^4 \}
 \label{formuleF}
\end{equation}

 \noindent When the external applied field  $h(x,t)$ is constant in  time,
 the invariant  measure   associated with Eq.~(\ref{modelA})
 is the equilibrium Gibbs-Boltzmann  distribution:
\begin{eqnarray}
  { P_{eq}}[\phi] = \frac{e^{-\beta {\mathcal U}[\phi] }}{Z[\beta, h]}
 \, \,  \hbox{ with } \, \,
  Z[\beta, h] = \int {\mathcal D}\phi
\   e^{-\beta{\mathcal U}[\phi] }
 \label{mesureinv}
\end{eqnarray}

  We  now consider  the case where the  applied field varies
 with time according to a  well-defined   protocol:
  For  $t\le 0$,
 we have $h(x,0) = h_0(x)$ and the system is in
 its stationary state;   for $t>0$, the external field  varies
 with time,
  reaches its    final  value $h_f(x)$
  after a finite time $t_f$, and remains constant  for $t \ge t_f$.
  The  values of the potential
  ${\mathcal U}$ for $t \le  0 $ and $t \ge t_f$ are
 denoted by  ${\mathcal U}_0$ and ${\mathcal U}_1$, respectively.

The  probability  ${\mathcal P}( \phi_1 |  \phi_0 )$ of observing
 the field  $\phi_1(x)$ at time $t_f$ starting from
 $\phi_0(x)$ at time $t=0$ is given by
  \begin{equation}
 {\mathcal P}( \phi_1 |  \phi_0 ) =
  \int {\mathcal D}\zeta
  ~{e}^{  -\frac{\beta}{4\Gamma_0}\int {\rm d}^d x  {\rm d}t \zeta^2}
    \delta \left(\phi(x,t_f)-  \phi_1(x) \right)
    \label{eq:defP}
\end{equation}
 The following identity is now substituted in Eq. (\ref{eq:defP})
  \begin{eqnarray}
   1=\int  {\mathcal D}\phi(x,t)  \delta
   \left( \dot\phi(x,t) + \Gamma_0 \frac {\delta {\mathcal U}}{\delta  \phi}
   -\zeta(x,t) \right) |\det{\bf M} |
   \nonumber
   \\=\int  {\mathcal D}\phi(x,t) {\mathcal D}\bar\phi(x,t)
   |\det{\bf M} |
   ~e^{-\int d^dx dt \bar\phi \{\dot\phi +
   \Gamma_0 \frac {\delta {\mathcal U}}{\delta  \phi}
   -\zeta \}} \,\,\,
  \label{eq:onedelta}
\end{eqnarray}
 where $\bar\phi(x,t)$ is  the response field   and ${\bf M}$ is the operator
 \begin{eqnarray}
  {\bf M} =  \frac{\partial}{\partial t} + \Gamma_0 \frac{\partial
  f(\phi(x,t),t)}{\partial \phi}
    \label{eq:defM}
 \end{eqnarray}
\noindent Integrating over the noise $\zeta(x,t)$, one finds
\cite{JZJ,MMJZJ}
 \begin{equation}
  {\mathcal P}( \phi_1 |  \phi_0 ) =
 \int_{\phi(x,0)= \phi_0(x)}^{\phi(x,t_f)= \phi_1(x)}
  {\mathcal D}\phi {\mathcal D}\bar\phi \,
  {e}^{ - \int {\rm d}^d x  {\rm d}t
  \Sigma(\phi,\dot\phi,\bar\phi)} \,
 \label{pathintforP}
\end{equation}
 where the dynamical action
$\Sigma$ is  given by
\begin{eqnarray}
 \Sigma(\phi,\dot\phi,\bar\phi)=  \Gamma_0  \bar\phi(
     \frac{\dot\phi}{\Gamma_0} +
 \frac{\delta {\mathcal U}}  {\delta  \phi}  -\frac{\bar\phi}{\beta})
  - \frac{ \Gamma_0 }{2}
 \frac{\delta^2{\mathcal U}}  {\delta \phi^2}
  \label{eq:action}
\end{eqnarray}
 the  last term being  the Jacobian of ${\bf M}$. We
 consider now a functional ${\mathcal O}[\phi]$
 that depends of the values of the field $\phi(x,t)$ for
 $0 \le t \le t_f$. The average of ${\mathcal O}[\phi]$
 with respect to the stationary
 initial ensemble and the stochastic evolution  between times
 $0$
 and $t_f$ is given by the path integral
\begin{eqnarray}
    \langle  {\mathcal O}  \rangle  =
\frac{1}{Z_0} \int {\mathcal D}\phi_0(x){\mathcal D}\phi_1(x)
{e}^{- \beta {\mathcal U}_0[\phi_0] }
~~{\mathcal I}\{ {\mathcal O} , \phi_1, \phi_0 \}
\label{meanvalueofO}
 \end{eqnarray}
\noindent where
\begin{eqnarray}
{\mathcal I}\{ {\mathcal O} , \phi_1, \phi_0 \}= \int_{\phi(x,0)=
\phi_0(x)}^{\phi(x,t_f)= \phi_1(x)}
  {\mathcal D}\phi(x,t) {\mathcal D}\bar\phi(x,t) \,
\nonumber \\
  {e}^{ - \int {\rm d}^d x  {\rm d}t
  ~\Sigma(\phi,\dot\phi,\bar\phi)}
  ~{\mathcal O}[\phi] \ \ .
 \label{defI}
 \end{eqnarray}

\noindent Under  a  change  of the integration variable $\bar\phi$
 in Eq.~(\ref{meanvalueofO}),
  \begin{eqnarray}
          \bar\phi(x,t) \rightarrow  -\bar\phi(x,t) +
  \beta \frac{\delta {\mathcal U}[\phi(x,t),t]}  {\delta  \phi(x,t)}
 \label{eq:shift}
\end{eqnarray}
the path integral measure is   invariant and $\Sigma$
  varies as
   \begin{eqnarray}
          \Sigma(\phi,\dot\phi, \bar\phi)    \rightarrow
  \Sigma(\phi,-\dot\phi, \bar\phi) +
  \beta {\dot\phi}  \frac{\delta {\mathcal U} [\phi(x,t),t]}
    {\delta  \phi(x,t)}
 \label{eq:variation}
\end{eqnarray}
 Writing the second  term  on the r.h.s. as
 $(\frac{d{\mathcal U}}{ dt} - \frac{ \partial{\mathcal U}}{\partial{t}})$,
gives
  \begin{eqnarray}
 \int  {\rm d}^d x  ~{\rm d}t \  {\dot\phi}
\ \frac{\delta {\mathcal U} [\phi(x,t),t]}  {\delta  \phi(x,t)} =
 {\mathcal U}_1[\phi_1] - {\mathcal U}_0[\phi_0] -  {\mathcal W}_J[\phi]
 \label{eq:variation2}
\end{eqnarray}
 The derivative $\partial
 {\mathcal U} / \partial {t}$ is related
 to Jarzynski's  work by
  \begin{equation}
  {\mathcal W}_J[\phi] =  \int_{0}^{t_f}
   {\rm d}t \, \frac{ \partial{\mathcal U}}{\partial{t}}
 =   -  
   \int {\rm d}^d x \, {\rm d}t  \,\,  \dot{h}(x,t)\phi(x,t)
 \label{eq:defWJ}
\end{equation}
 the last equality being  obtained from   Eq.~(\ref{def:Ut}).
The change of sign of the time derivative $\dot\phi$
  in  Eq.~(\ref{eq:variation})  is compensated  by the
 change of   variables  in the  path integral
 \begin{eqnarray}
   \left(\phi(x,t), \bar\phi(x,t)\right)
  \rightarrow  \left(\phi(x,t_f -t), \bar\phi(x, t_f -t)\right)
\label{eq:Tsym}
  \end{eqnarray}
  This time-reversal  transformation
  leaves the functional  measure invariant and restores $\Sigma$
 to its original form but with a {\it time-reversed}  protocol
 for the external applied field $h(x,t) \rightarrow h(x,t_f -t)$.
Performing the above change of variables
 (\ref{eq:shift}) and~(\ref{eq:Tsym})
 in
 Eq. (\ref{defI}) and using
  Eqs. (\ref{eq:variation}) ~
  and~(\ref{eq:variation2}),
 we find
\begin{eqnarray}
 {\mathcal I}\{ {\mathcal O} , \phi_1, \phi_0 \}=
 {e}^{ \beta ({\mathcal U}_0[\phi_0]-
 {\mathcal U}_1[\phi_1])}
 {\mathcal I}\{ {e}^{-\beta {\mathcal W}_J}\hat{\mathcal O} ,
  \phi_0,
 \phi_1\}_R \hskip .5cm
 \label{Crooks0}
\end{eqnarray}
 The subscript $R$ denotes a
 time-reversed protocol
  and the time-reversed  $\hat{\mathcal O}[\phi]$ is
   equal to
    ${\mathcal O}[\phi(x,t_f -t)]$.
Inserting this identity in Eqs.~(\ref{meanvalueofO} - \ref{defI})
, gives
\begin{eqnarray}
   {  \langle  {\mathcal O}  \rangle} &=&
  \frac{1}{Z_0} \int {\mathcal D}\phi_0(x) {\mathcal D}\phi_1(x)
 {e}^{ -\beta{\mathcal U}_1[\phi_1]}
   ~~{\mathcal I}\{ {e}^{-\beta {\mathcal W}_J}\hat{\mathcal O} ,
  \phi_0,
 \phi_1\}_R
       \nonumber  \\
 &=&  \frac{Z_1}{Z_0}
  {  \langle  \hat{\mathcal O}
  {e}^{-\beta {\mathcal W}_J} \rangle_R} = {e}^{-\beta \Delta F}
  {  \langle  \hat{\mathcal O}
  {e}^{-\beta {\mathcal W}_J} \rangle_R}
  \end{eqnarray}
  where $\Delta F$ is the free energy difference
  between the final and the initial states.
   Finally,   we redefine   ${\mathcal O}$
 as   ${\mathcal O} {e}^{- \beta W_J}$.
  Recalling that the work ${\mathcal W}_J$
 is odd under time-reversal,  we deduce  from the last equation  that
 \begin{equation}
    {\langle  {\mathcal O} {e}^{-\beta {\mathcal W}_J}   \rangle}
 = {e}^{-\beta \Delta F}   {  \langle  \hat{\mathcal O}  \rangle_R}
 \label{OhatO}
 \end{equation}
  When   ${\mathcal O} = 1$, we obtain
 Jarzynski's  theorem
\begin{equation}
  \langle {e}^{- \beta W_J}   \rangle  =  {e}^{-\beta \Delta F}
 \label{eq:Jarzynski}
\end{equation}
   Taking  ${\mathcal O}  =  {e}^{(\beta-\lambda){\mathcal W}_J}$,
 where $\lambda$ is an arbitrary real parameter,
  we derive   the following symmetry  property
\begin{equation}
       \langle {e}^{- \lambda W_J}   \rangle  =  {e}^{-\beta \Delta F}
      \langle {e}^{(\lambda -\beta) W_J}   \rangle_R
      \label{eq:Laplace}
\end{equation}
 and a Laplace transform
 leads  to  Crooks relation
  \cite{crooks1,crooks2}:
\begin{equation}
   \frac{ {\mathcal P}_F(W) }{ {\mathcal P}_R(-W)} =
{e}^{ \beta (W - \Delta F)  }
 \label{eq:Crooks}
\end{equation}
 where  ${\mathcal P}_F$ and  ${\mathcal P}_R$ represent
 the   probability distribution functions  of the work
 for the forward  and the  reverse processes, respectively.
 We emphasize that our  proof
  of Crooks and Jarzynski  identities  is  based  on  invariance
 properties of the path integral
  and does not involve any  a priori  thermodynamic definition of heat
 and work. The  expression~(\ref{eq:defWJ}) for  the
  Jarzynski work
  appears as a natural outcome of  this  invariance.

  The identity~(\ref{OhatO}), which is at the core of the
  work fluctuation relations, is valid for any choice of  the
  external field protocol. The free energy variation is  a function  only
  of the extremal values of the applied field  at $t_0 =0$ and $t = t_f$
  and is independent of  the values  at intermediate
  times. Functional derivatives  of Eq.~(\ref{eq:Jarzynski})
  with respect to  $h(x,t)$ at an intermediate time $ t_0 < t < t_f,$
  and at position $x$,  results in new  identities
\begin{equation}
    \langle \big( \bar\phi(x,t) - \frac{\beta}{\Gamma_0}
  \dot\phi(x,t) \big) {e}^{-\beta W_J}  \rangle
   =  0
 \label{eq:deriveejarz}
\end{equation}
 The $n$-th functional derivative of Eq.~(\ref{eq:Jarzynski})
  at intermediate times $t_1,\ldots t_n$ and positions
   $x_1,\ldots x_n$, gives the identity
\begin{equation}
    \langle    {e}^{-\beta W_J} \prod_{i=1}^n
  \big(\bar\phi(x_i,t_i) - \frac{\beta}{\Gamma_0}
  \dot\phi(x_i,t_i) \big)   \rangle
   =  0
 \label{eq:deriveeniemejarz}
\end{equation}
 Similarly, the  functional derivative  of Eq.~(\ref{OhatO}) leads to
\begin{equation}
   \langle (\bar\phi(x,t) - \frac{\beta}{\Gamma_0}
 \dot\phi(x,t)) {\mathcal O}  {e}^{- \beta W_J}  \rangle =
    {e}^{-\beta \Delta F} \langle{\hat{\bar \phi} (x,t)}  {\hat {\mathcal O}}  \rangle_R
 \label{eq:derivee1}
\end{equation}
 Eq.(\ref{eq:deriveejarz})~follows by choosing
  ${\mathcal O}=\hat{\mathcal O}=\hat{\bf 1}$
  since {\cite{footnote1}}
   $\langle {\bar\phi}\rangle=0$.
 For  the special  case ${\mathcal O}[\phi] =
 \phi(x',t')$, we obtain
a generalization of the fluctuation-dissipation theorem \begin{eqnarray}
 \label{eq:GenFDT}
&&
\frac {\delta \langle \phi(x',t') {e}^{-\beta W_J} \rangle }{\delta h_1 (x,t)} \Big |_{h_1=0}  -\beta  \langle \dot\phi(x,t))  \phi(x',t')   {e}^{-\beta W_J} \rangle  \nonumber
= \\
&&{e}^{-\beta \Delta F} \frac {\delta  \langle \hat \phi(x',t') \rangle_R}{\delta \hat h_1 (x,t)}\Big |_{h_1=0} 
\end{eqnarray}
where $h_1(x,t)$ is a small perturbation that drives the system out of the protocol $h(x,t)$. Note that $h_1$ does not enter the definition of the Jarzynski work.
This new fluctuation-dissipation theorem relates out of equilibrium response functions to derivatives of correlation functions and could be verified experimentally, for example in single molecule pulling experiments (this corresponds to the case where the field $\phi$ does not depend on space).
 For   a system at  thermodynamic  equilibrium
  with constant  external field ({\it i.e.},
 $ W_J = \Delta F =0$) and stationary
 correlations,   this equation  reduces
   to  the standard fluctuation-dissipation relation  \cite{chaturvedi}

 Identities between correlators such as
   Eqs.~(\ref{eq:deriveejarz})-(\ref{eq:derivee1})   suggest  the existence
 of an underlying continuous  symmetry of the  system.
We first extend the integration range of the path integral in
Eq.~(\ref{meanvalueofO}) over the range  $-\infty < t < +\infty$,
using the following properties of the
   probability distribution:
 \begin{eqnarray}
\frac{1}{Z_0}
  {e}^{- \beta {\mathcal U}_0[\phi_0]}=
      \lim_{T\to -\infty} P(\phi_0  | \phi_{T}) \hskip 1cm
 \label{ergodicity} \\
       1 =  \int {\mathcal D}\phi(x,T)  P(\phi_T  | \phi_{t_1})
 \,\,\,\, \hbox{ for  } \,\,\,\,  T > t_1  \label{normalization}
 \end{eqnarray}
 The first property assumes ergodicity and the latter is
 normalization. Inserting into these equations the path integral
 representation, Eq.(\ref{pathintforP}) ,
  of  $ P(\phi'  | \phi'')$,
 ~Eq.(\ref{meanvalueofO}) is rewritten as
\begin{equation}
    \langle  {\mathcal O}  \rangle  =
   \int {\mathcal D}\phi(x,t){\mathcal D}\bar\phi(x,t)
  ~{e}^{ - \int {\rm d}^d x  {\rm d}t
  \Sigma(\phi,\dot\phi,\bar\phi)}
  {\mathcal O}[\phi]   \, \,\,\,\, \\
  \label{eq:extendedpathint}
\end{equation}
 where $\phi(x,t)$ and $\bar\phi(x,t)$ are  integrated with t ranging now from  $
  -\infty$ ~to~ $\infty$.

  \noindent To uncover the above mentioned
  hidden symmetry, in addition to
 the original field $\phi(x,t)$ and  the response field
 $\bar\phi(x,t)$,   we introduce  two auxiliary
 anti-commuting  Grassmannian fields $c(x,t)$ and $\bar{c}(x,t)$
  that allow us to express  the Jacobian of ${\bf M}$, defined in
 Eq.~(\ref{eq:defM}),   as a functional
 integral \cite{gozzi,chaturvedi}.
Assuming that   ${\mathcal O}$
  differs from the identity only for $0 \le t \le  t_f ~,$
 the mean value~  of  ${\mathcal O}$ in Eq.(\ref{meanvalueofO}) can be
 rewritten as
\begin{eqnarray}
    \langle  {\mathcal O}  \rangle  =
   \int {\mathcal D}\phi {\mathcal D}\bar\phi
  {\mathcal D}c {\mathcal D}\bar{c} \,\,
  ~{e}^{ - \int {\rm d}^d x  {\rm d}t
  {\bf \Sigma}(\phi,\bar\phi,c,\bar{c})}
  ~~{\mathcal O}[\phi] \, \hskip 1cm ,
  \label{eq:pathint3}
\end{eqnarray}
\vskip -.5cm
\begin{eqnarray} {\rm where~~~}
 {\bf \Sigma}(\phi,\bar\phi,c,\bar{c})=  \Gamma_0  \bar\phi(
     \frac{\dot\phi}{\Gamma_0} +
 \frac{\delta {\mathcal U}}  {\delta  \phi}  -\frac{\bar\phi}{\beta})
  - c{\bf M} \bar{c}   \,\,
 \label{eq:susyaction}
\end{eqnarray}
 \vskip -.1cm
\noindent with ${\bf M}$ given in Eq.~(\ref{eq:defM}).

Consider now   the    infinitesimal   transformation that mixes
 ordinary fields  with  Grassmannian  fields:
\vskip -.6cm
\begin{eqnarray}
   \delta \phi(x,t) &=&  c(x,t) \bar\epsilon   {\hskip 0.7cm}
 \delta \bar  \phi(x,t) =  \frac{\beta}{\Gamma_0}\,
\dot{c}(x,t) \bar\epsilon  \label{eq:BRSTsymmetry}   \\
   \delta  c(x,t) &=&  0    {\hskip 0.5cm}    \delta \bar{c}(x,t) =
  \left(\bar\phi(x,t) - \frac{\beta}{\Gamma_0}
 \dot\phi(x,t)\right)\bar\epsilon
 \label{eq:SUSY} \nonumber
\end{eqnarray}
   $\bar\epsilon$ being  a time-independent  infinitesimal Grassmannian field.
  The variation of ${\bf \Sigma}(\phi,\bar\phi,c,\bar{c})$~in
  Eq.(\ref{eq:susyaction}) under
 the transformation~of Eq.(\ref{eq:BRSTsymmetry}) gives
 \begin{eqnarray}
 \delta{\bf \Sigma}(\phi,\bar\phi,c,\bar{c}) &=&
 \frac{d {\mathcal A} }{dt}  - \beta \,
\frac{\partial}{\partial t} \left(  \frac{\delta {\mathcal U}}
{\delta  \phi} \right)
  c(x,t)   \,  \bar\epsilon
 \label{eq:noncovariance}
 \end{eqnarray}
  \vskip -.3cm
 with the total derivative term
  \begin{eqnarray}
 {\mathcal A}  &=&
   \beta  \,  \Big(  \frac{\dot\phi}{\Gamma_0} +
 \frac{\delta {\mathcal U}}  {\delta  \phi}  -\frac{\bar\phi}{\beta} \Big)
   c \,  \bar\epsilon   \label{eq:defA}
\end{eqnarray}
\vskip -.3cm
   \noindent If the potential ${\mathcal U}$ is independent of time,   the
 variation of  ${\bf \Sigma}$  under  the  supersymmetric
  transformation~(\ref{eq:BRSTsymmetry})
  is a total  time derivative that  does not modify the
 action.  The   supersymmetry in Eq.({\ref{eq:BRSTsymmetry}})~which is in fact
 a generalization of supersymmetric quantum mechanics {\cite{msr,JZJ}},  reflects
 the time reversal  invariance of
  Model A  in absence of external field and
  allows to prove the fluctuation-dissipation theorem
  \cite{gozzi, chaturvedi}.

 \noindent When the  potential  ${\mathcal U}(\phi,t)$ depends explicitly on  time,
 supersymmetry invariance is broken: the last term in
 Eq.~(\ref{eq:noncovariance})  breaks the invariance. It can  be written as
 \begin{equation}
   \beta \frac{\delta^2{\mathcal U}}{\delta \phi \partial t} c \,
      \bar\epsilon =  \beta
  \frac{\delta^2{\mathcal U}}{\delta \phi \partial t} \delta \phi
 = \delta\left( \beta \frac{\partial{\mathcal U}}{\partial t}\right)
 \label{contreterme}
 \end{equation}
 and  can   be interpreted as the variation
 of a function. Hence, the   modified
 $\Sigma_J$,
 defined as
 \vskip -.5cm
  \begin{equation}
    \Sigma_J = \Sigma + \beta \frac{\partial{\mathcal U}}{\partial t}
 \label{def:SigmaJ}
 \end{equation}
 and  obtained by   adding  the Jarzynski work~(\ref{eq:defWJ})
 to the initial action, {\it is invariant}  under  the
  supersymmetric transformation ~(\ref{eq:BRSTsymmetry})  up
 to a  total derivative term:
  $
   \delta  \Sigma_J =  \frac{d {\mathcal A} }{dt}
 $.
The boundary terms at $t=\pm\infty$ are, conventionally, assumed
to vanish.

 We now show that the  supersymmetric  invariance  of
 $\Sigma_J$ implies
 the  correlator identities~(\ref{eq:deriveeniemejarz}).
   Introducing  a four-component source
  $
  (H, \bar{H},  \bar{L}, L)$ ,
   we define  the  generating function
 \begin{eqnarray}
   Z(H, \bar{H},  \bar{L}, L)  =
  \int  {\mathcal D}\phi {\mathcal D}\bar\phi
  {\mathcal D}c {\mathcal D}\bar{c}
  \hskip 2cm
  \nonumber \\
 \exp
 \left(
  \int {\rm d}^d x      {\rm d}t
  \left( -{\bf \Sigma}_J(\phi ,\bar\phi,
 c ,\bar{c}) +
   \bar{H} \phi +  H \bar\phi
     + \bar{L} c  + L \bar{c}\right)
     \right)
      \nonumber
 \label{def:Z}
   \end{eqnarray}
   Making the transformation~(\ref{eq:BRSTsymmetry}) in  $Z(H, \bar{H},  \bar{L}, L)$
   and using   the  supersymmetric
  invariance~
  of ${\bf \Sigma}_J$~ in Eq.(\ref{def:SigmaJ}),
   we deduce as in \cite{chaturvedi}
  the   Ward-Takahashi  identity:
 \begin{equation}
    \int  \left(  
  \frac{\beta}{\Gamma_0}
 H \frac{{\rm d}}{{\rm d}t} \frac{\delta Z}{\delta \bar{L}}
 +{L} \left(  \frac{\delta Z}{\delta H} - \frac{\beta}{\Gamma_0}
    \frac{{\rm d}}{{\rm d}t}  \frac{\delta Z}{\delta \bar{H}} \right)
 +  \bar{H} \frac{\delta Z}{\delta \bar{L} } \right)
       = 0
\label{Ward2}
 \end{equation}
 By applying  to the   Ward-Takahashi identity  the  operator
$  \frac{\delta}{\delta L(x,t)}  \,
 \prod_{i=1}^n  \left(  \frac{\delta}{\delta H(x_i,t_i)}
    -\frac{\beta}{\Gamma_0}  \frac{{\rm d}}{{\rm d}t_i}
  \frac{\delta}{\delta \bar{H}(x_i,t_i) } \right)   $
 and setting the source field $H, \bar{H},  \bar{L}, L$ to zero,
  we obtain
  Eqs.~(\ref{eq:deriveejarz}) and (\ref{eq:deriveeniemejarz}).
This lead to Jarzynski's  equality~(\ref{eq:Jarzynski}).
   Replacing
   $h(x, t)$   by  $h(x, \alpha t)$
  for any  $\alpha > 0$, we obtain
 \begin{eqnarray}
  \frac{d \langle e^{- \beta W_J} \rangle }
  {d \alpha} =
   \int {\rm d}^d x  \, {\rm d}t  ~t~     \dot{h}(x,\alpha t)   \langle
 ( \bar\phi(x,t) -  \frac{\beta}{\Gamma_0} \dot\phi(x,t) )
 {e}^{-\beta  W_J}  \rangle  \nonumber
 \end{eqnarray}
  Using Eq.(\ref{eq:deriveejarz}) that can be obtained
 from  (\ref{Ward2}),  we get
 $ {d \langle {e}^{- \beta W_J}\rangle }/{d \alpha} =  0$,
 which  means that the  value of
 $\langle {e}^{- \beta W_J} \rangle$ does not
 depend on $\alpha$. Hence,  this value
 is the same as that of  the  quasi-static limit $\alpha \to 0$,
  and  is  given by  $\exp(-\Delta F)$.  Jarzynski's identity
 is thus obtained  as a consequence of supersymmetry.

 The response-field  technique in Eqs.({\ref{pathintforP}}-{\ref{defI}})
 that we have used to derive
  nonequilibrium work theorems  for
  Model~A
  can be extended to multi-component
  fields and  to  other stochastic models  such as model B.
  It also can be extended to systems with correlated  noise {\cite{Adhar}}
  replacing the Gaussian measure in the RHS of Eq.({\ref{eq:defP}}) by
\begin{equation}
  \int {\mathcal D}\zeta
  ~{e}^{  -\frac{1}{2}\int
  {\rm d}^d x  ~{\rm d}t   ~{\rm d}^d y ~{\rm d}t'
  ~\zeta(x,t)~\Delta^{-1}(x,t;y,t')~\zeta(y,t')}
    \label{eq:correlatedNoise}
\end{equation}
 where $ \Delta(x,t;y,t')$ is the two point correlation function.

 We  have obtained  correlators
  identities involving  an  arbitrary field-operator
 and also  a  generalization of the
 fluctuation-dissipation relation that remains valid far from equilibrium.
 The supersymmetric invariance of the time
 independent Langevin equation  breaks down
when  the potential varies according to a time-dependent protocol.
We have shown
 that the supersymmetry in Eq.({\ref{eq:SUSY}}) is restored by adding  to
 the action  a  counter-term
 which is precisely  the  Jarzynski work $\beta{\mathcal W[\phi]}_J$.
  Furthermore, we proved  that
    the associated  supersymmetric  Ward Identity implies
   Jarzynski's theorem.
  Supersymmetry enforces   the exactness of
 the quasi-static limit even for processes that have a finite duration
 and that  bring the system arbitrarily  far from equilibrium.
 A hidden supersymmetry \cite{gozzi2} is  also present in classical
 Hamiltonian systems  for which Jarzynski's equality
 was initially proved. We finally remark
 that supersymmetry may also be a useful tool when applied to
 the fluctuation theorem for stochastic dynamics \cite{Kurchan}.


\vskip -.5cm
\begin{thebibliography}{article}
 \bibitem{jarzynskiPRL} C.~Jarzynski,  Phys. Rev. Lett. {\bf 78}, 2690 (1997).
\bibitem{jarzynskiPRE}  C.~Jarzynski,   Phys. Rev. E {\bf 56}, 5018 (1997).
 \bibitem{crooks1}  G.~E.~Crooks, J. Stat. Phys. {\bf 90}, 1481 (1998);
 Phys. Rev. E {\bf 60}, 2721 (1999).
\bibitem{crooks2}  G.~E.~Crooks,    Phys. Rev. E  {\bf 61},  2361 (2000).
   \bibitem{jarmazonka} O. Mazonka and  C.~Jarzynski, arXiv:cond-mat/9912121.
\bibitem{lua} R.~C.~Lua and A.~Y.~Grosberg, J.~Phys.Chem. B {\bf 109},
 6805 (2005).
 \bibitem{vandenbroeck} I.~Bena, C.~Van~den~Broeck and R.~Kawai,
 Euro. Phys. Lett. {\bf 71}, 879 (2005).
\bibitem{hummer} G.~Hummer and A.~Szabo, Proc. Nat. Acad. Sci. USA,
  {\bf 98}, 3658 (2001).
\bibitem{liphardt} J.~Liphardt, S.~Dumont,  S.~B.~Smith, I.~Tinoco
 and C.~Bustamante, Science,  {\bf 296}, 1832 (2002).
\bibitem{ritort} D. Collin, F.~Ritort, C.~Jarzynski, S.~B.~Smith,
 I.~Tinoco and C.~Bustamante, Nature {\bf 437} 231 (2005).
 \bibitem{douarche}   F.~Douarche, S.~Ciliberto and A.~Petrosyan,
J. Stat. Mech.: Theor. Exp. P09011 (2005).
\bibitem{ritortrev} F.~Ritort, Sem. Poincar\'e {\bf 2}, 193 (2003);
cond-mat/0401311.
  \bibitem{hohenberg}  P.~C.~Hohenberg  and   B.~I.~Halperin,
   Rev. Mod. Phys.  {\bf 49},  435 (1977).
\bibitem{abishek1}  O.~Narayan and  A.~Dhar, J. Phys. A: Math. Gen.
  {\bf 37}, 63 (2004).
\bibitem{jarzynski4} V.~Y.~Chernyak, M.~Chertkov and C.~Jarzynski,
   Phys. Rev. E  {\bf 71}, 025102(R) (2005);
  arXiv:cond-mat/0605547.
 \bibitem{chetrite} R.~Chetrite and K.~Gaw\c{e}dzki,  arXiv:math-phy/07072725
\bibitem{msr} P.~C.~Martin, E.~D.~Siggia and H.~A.~Rose,
  Phys. Rev. A {\bf 8}, 423 (1973);  C.~de~Dominicis and L.~Peliti,
  Phys. Rev. B  {\bf 18}, 353 (1978).
 \bibitem{tsvelik}  M.~V.~Feigel'man and A.~M.~Tsvelik,
  Sov. Phys. JETP  {\bf 56}, 823 (1982);
 Phys. Lett. A. {\bf 95 A}, 469 (1983).
\bibitem{gozzi} E.~Gozzi, Phys. Rev. D {\bf 30}, 1218  (1984).
 \bibitem{chaturvedi} S.~Chaturvedi, A.~K.~Kapoor
 and S.~Srinivasan, Z. Phys. B   {\bf 57}, 249  (1984).
\bibitem{JZJ} J.~Zinn-Justin, Quantum Field Theory and Critical
 Phenomena, Fourth Edition (Clarendon Press, Oxford 2002).
\bibitem{MMJZJ} M.~Moshe and  J.~Zinn-Justin, Phys. Rep.
 {\bf 385}, 69 (2003).
 \bibitem{footnote1}Using the functional derivative
 of Eq.({\ref{meanvalueofO}}) with respect to $h(x,t)$ for  ${\mathcal O}={\bf
 1}$, one finds: $\langle \bar\phi \rangle = \langle \frac{\dot\phi}{\Gamma_0} +
 \frac{\delta {\mathcal U}}  {\delta  \phi} \rangle = 0.$
\bibitem{Adhar} T. Mai and A. Dhar, Phys. Rev. E {\bf 75}, 061101  (2007)
 \bibitem{gozzi2} E. Gozzi, M.~Reuter and W.~D.~Thacker,
 Phys. Rev. D {\bf 40}, 3363  (1989).
 \bibitem{Kurchan} J. Kurchan, ~J. Phys. A: Math. Gen. {\bf 31},
 3719 (1998).
\end{thebibliography}
\end{document}